\documentclass[12pt,letterpaper]{article}
\usepackage{osajnl2}
\usepackage[draft]{hyperref} 
\newcommand\beq{\begin{equation}}
\newcommand\eeq{\end{equation}}

\begin{document}

\title{Power Scattering and Absorption Mediated by Cloak/Anti-Cloak Interactions:
A Transformation-Optics Route towards Invisible Sensors}

\author{Giuseppe Castaldi,$^1$ Ilaria Gallina,$^1$ Vincenzo Galdi,$^{1,*}$ Andrea Al\`u,$^2$ and Nader Engheta$^3$}
\address{$^1$Waves Group, Department of Engineering, University of Sannio,
Corso Garibaldi 107, I-82100 Benevento, Italy}
\address{$^2$ Department of Electrical and Computer Engineering, The University of Texas at Austin, Austin, TX 78712, USA}
\address{$^3$ Department of Electrical and Systems Engineering, University of Pennsylvania, Philadelphia, PA 19104, USA}

\address{$^*$Corresponding author: vgaldi@unisannio.it}




\begin{abstract}
The suggestive idea of ``cloaking'' an electromagnetic sensor, i.e., strongly reducing its visibility
(scattering) while maintaining its field-sensing (absorption) capabilities, has recently been
proposed in the literature, based on scattering-cancellation, Fano-resonance, or transformation-optics
approaches.
In this paper, we explore an alternative, transformation-optics-based route, which relies on the
recently-introduced concept of ``anti-cloaking.'' More specifically, our proposed approach relies on
a suitable tailoring of the competing cloaking and anti-cloaking mechanisms, interacting in a two-dimensional
cylindrical scenario. Via analytical and parametric studies, we illustrate the underlying
phenomenology, identify the critical design parameters, and address the relevant optimality and
tradeoff issues, taking also into account the effect of material losses. Our results confirm the envisaged
potentials of the proposed transformation-optics approach as an attractive alternative route to sensor
cloaking.
\end{abstract}


\section{Introduction}

During the past few years, there has been a growing interest in the development of strategies and
devices to achieve {\em invisibility} of objects to waves of various nature (electromagnetic, acoustic,
elastic, matter). With special reference to the electromagnetic (EM) case, among the most
prominent approaches to passive ``invisibility cloaking,'' it is worth mentioning those based on
scattering cancellation \cite{Alu1,Silveirinha},  anomalous localized resonances \cite{Milton},  transformation optics (TO) \cite{Leonhardt,Pendry}, inverse
design of scattering optical elements \cite{Hakansson}, transmission-line networks \cite{Alitalo1},  and tapered waveguides \cite{Smolyaninov2}. The reader is also referred to \cite{Schurig,Smolyaninov1,Alitalo2,Alu3} for related experimental studies at microwave and optical
frequencies.

In the TO approach to cloaking, of particular interest for what follows, the rerouting of the
impinging energy around the concealed object is first designed in a fictitious curved-coordinate
space containing a ``hole,'' and it is subsequently translated into a conventionally flat, Cartesian
space, filled by an anisotropic and spatially inhomogeneous ``transformation medium'' which
embeds the coordinate-mapping effects \cite{Leonhardt,Pendry,Shalaev,Leonhardt2}. Starting from this basic concept, a variety of
extensions and twists have been proposed, including the carpet cloak \cite{Li,Liu,Gabrielli,Valentine,Ergin,Ma1}, open cloak \cite{Ma}, cloak at a distance \cite{Lai}, anti-cloak \cite{Chen,Castaldi},  superscatterer \cite{Yang} and superabsorber \cite{Ng}, invisible gateway \cite{Chen1},  as well as the broad framework of ``illusion optics'' \cite{Lai1}. 

In connection with the anti-cloaking concept, directly related to the subject of this paper, it was
shown in \cite{Chen} that, in spite of the generally assumed {\em impenetrability} of an ideal cylindrical cloak
shell \cite{Ruan1,Chen3,Chen4} (implying a lossless, anisotropic, spatially inhomogeneous medium, with extreme values of the
constitutive parameters ranging from zero to infinity), field penetration in the cloaked region may
actually occur in the presence of a suitably designed complementary {\em double-negative} (DNG) transformation
cylindrical shell of ``undoing'' (partially or completely) the cloak transformation. Expanding upon
this concept, in \cite{Castaldi}, we studied a more general scenario featuring a cloak and an anti-cloak
separated by a vacuum layer, and showed that: {\em i)} anti-cloaking-type effects and interesting field tunneling
phenomena may also be achieved with {\em double-positive} (DPS) or {\em single-negative} (SNG)
transformation media, thereby relaxing some of the practical feasibility limitations; {\em ii)} besides
the originally proposed ``cloaking countermeasure'' application in \cite{Chen}, alternative application
scenarios may be envisaged for which a (multiply-connected) region of space may be cloaked,
while maintaining the capability of somehow ``sensing'' the outside field from the inside. A similar
idea, within the suggestive framework of ``sensor cloaking,'' was recently explored in \cite{Alu2} (via a
scattering-cancellation approach), in \cite{Ruan} (via a Fano-resonance-type approach), and in \cite{Greenleaf}
(via a TO-based approach).

In this paper, we further elaborate on the above concepts, exploring more in detail the possibility to
apply them to sensor-cloaking. In particular, for the same two-dimensional (2-D) scenario
considered in \cite{Castaldi}, consisting of a metamaterial cylindrical target surrounded by a cloak and an
anti-cloak separated by a vacuum gap, we present some analytical results and parametric studies
aimed at gaining a deeper insight in the tradeoff between power absorption and scattering for a
slightly lossy target (mimicking the sensor loading effects). In this framework, we show that the
cloak/anti-cloak interactions can be tailored so as to allow sensible power absorption by the target,
while still guaranteeing very weak overall scattering, thereby indicating an alternative TO-based
route to sensor cloaking.

Accordingly, the rest of the paper is organized as follows. In Sec. \ref{Sec:statement}, we introduce the problem
geometry and formulation. In Sec. \ref{Sec:analytica}, we outline the general analytical solution and the relevant
derivations. In Sec. \ref{Sec:parametric}, we illustrate and discuss some representative results from a comprehensive
parametric study, and, in Sec. \ref{Sec:implementation}, we address some implementation-related issues. Finally, in Sec. \ref{Sec:conclusions}, we provide some brief concluding remarks and hints for future
research.

\section{Problem geometry and statement}
\label{Sec:statement}
\begin{figure}
\begin{center}
\includegraphics[width=6cm]{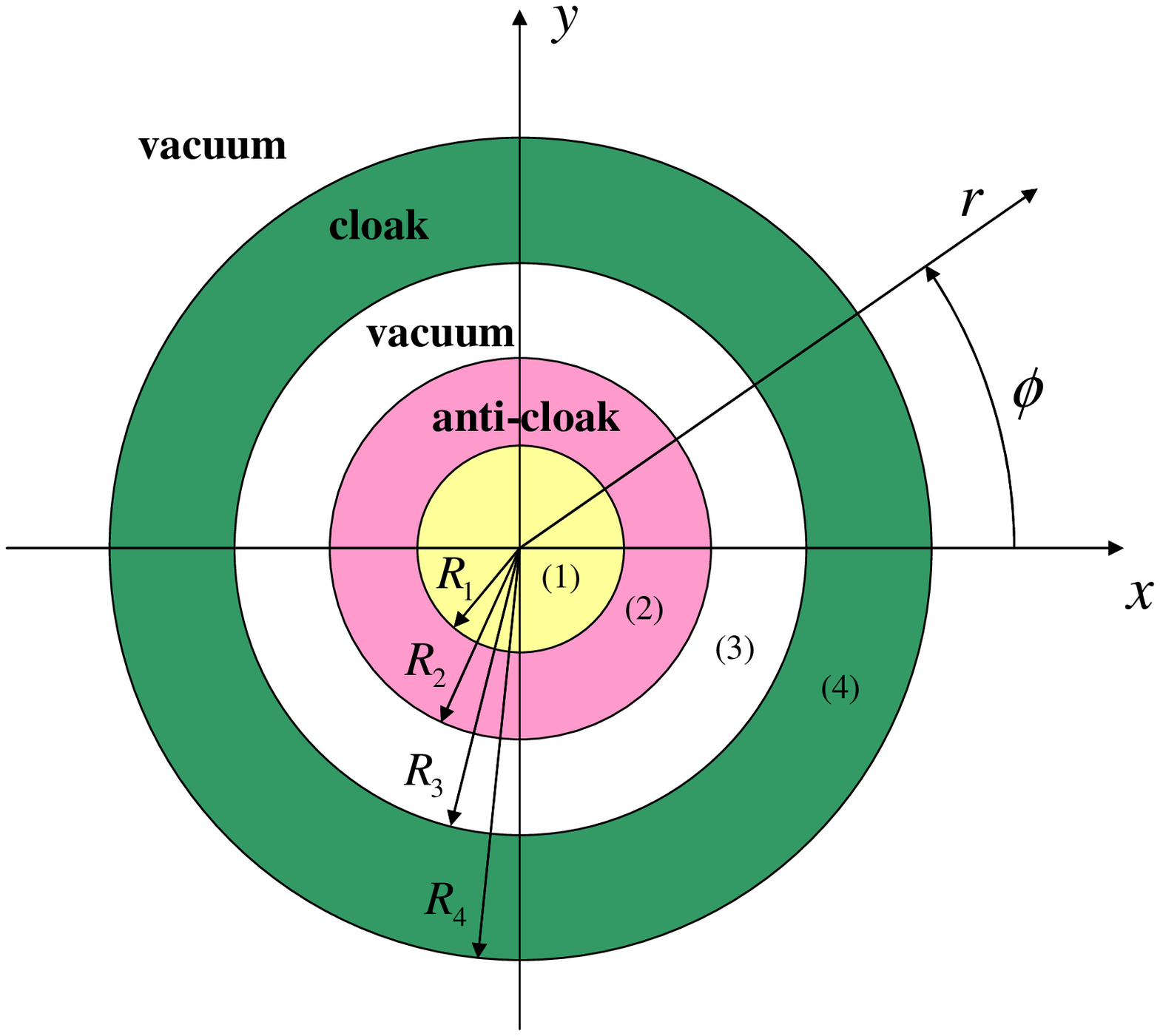}
\end{center}
\caption{(Color online) Problem geometry: A homogeneous metamaterial circular-cylindrical target of radius $R_1$ and constitutive parameters $\varepsilon_1$ and $\mu_1$, surrounded by a cloak and
anti-cloak cylindrical shells separated by a vacuum gap, in the
actual physical space $(x,y,z)$ [and associated cylindrical coordinate system $(r,\phi,z)$].}
\label{Figure1}
\end{figure}

The scenario of interest is the same as the one considered in \cite{Castaldi} and, as illustrated in Fig. \ref{Figure1}, entails a four-layer cylindrical configuration of radii $R_\nu, \nu=1,...,4$ in the physical space ($x,y,z$). More specifically, such configuration features an innermost metamaterial cylindrical target of radius $R_1$ (with electric permittivity $\varepsilon_1$ and magnetic permeability $\mu_1$) surrounded by two transformation-medium layers separated by a vacuum gap, and is obtained via a piecewise-linear radial coordinate mapping,
\beq
r' =f(r)= \left\{ \begin{array}{l}
r,~~ r < R_1 ,~~r > R_4 , \\ 
~\\
 R_1 \left( \displaystyle{{\frac{{R_2  + \Delta _2  - r}}{{R_2  + \Delta _2  - R_1 }}}} \right),~~ R_1  < r < R_2 , \\ 
~\\
 R_4 \left( \displaystyle{{\frac{{r - R_3  + \Delta _3 }}{{R_4  - R_3  + \Delta _3 }}}} \right),~~ R_3  < r < R_4 , \\ 
 \end{array} \right.
\label{eq:CT}
\eeq
from an auxiliary fictitious space ($x',y',z'$) featuring a homogeneous circular cylinder with constitutive parameters $\varepsilon_1$ and $\mu_1$, and radius $R_2>R_1$. The analytical expressions of the constitutive tensors characterizing the cloak and anti-cloak transformation media are systematically derived from the (derivative of) the mapping function $f$ in (\ref{eq:CT}), and are not reported here for brevity (see \cite{Castaldi} for details).
It can be observed that, letting to zero the small parameters $\Delta_2$ and $\Delta_3$ [which parameterize the vanishing behavior of the coordinate transformation in (\ref{eq:CT})], the outermost layer $R_3<r<R_4$ reduces to the standard invisibility cloak \cite{Schurig} 
while the internal layer $R_1<r<R_2$ reduces to an anti-cloak \cite{Chen} (with constitutive parameters opposite in sign to those of the inner target). It was shown in \cite{Castaldi} that the two competing cloak/anti-cloak effects can be tailored so as to create an {\em effectively cloaked} region in the vacuum gap $R_2<r<R_3$, while still being able to restore a non-negligible field in the (lossless) inner cylinder. Such restored field may be interestingly used for sensing purposes, suggesting a novel route towards the sensor-cloaking idea. Therefore, it is of interest to study the scattering and absorption properties of the above configuration in the presence of a {\em slightly lossy} target in the inner region ($r<R_1$), and explore the role of the various parameters available.

\section{Analytical results}
\label{Sec:analytica}

\subsection{General solution}
Assuming time-harmonic $[\exp(-i\omega t)]$  transverse-magnetic (TM) plane-wave excitation (with unit-amplitude $z$-directed magnetic field) impinging from the positive $x$-direction, the relevant magnetic field in the four-layer configuration of Fig. \ref{Figure1} can be computed via our previously derived Fourier-Bessel expansion \cite{Castaldi}, based on the approaches originally proposed in \cite{Ruan1,Chen2,Chen3,Luo1},
\begin{eqnarray}
H_z\!\left({r,\phi}\right)\!\!&=&\!\!\!\!\!
\sum\limits_{n=-\infty}^\infty
\!\!\!\left\{\!{\left({a_n^{(\nu)}\!+\!\delta _{\nu 5}i^n}\!\right)\!J_n\left[{g\left(r \right)}\right]\!+\!b_n^{(\nu)}Y_n\left[{g\left(r \right)}\right]}\!\right\}\nonumber\\ 
&\times & \exp\left({in\phi}\right),~~R_{\nu-1}\!\!<r\!\!<R_\nu,~~\nu\!=\!1,..,5.
\label{eq:BFE}
\end{eqnarray}
In (\ref{eq:BFE}), $J_n$ and $Y_n$ denote the $n$th-order Bessel functions of the first and second kind \cite{Abramowitz}, respectively, $R_0 = 0$ and $R_5=\infty$ are ``dummy'' parameters introduced for notational convenience, $\delta_{pq}$ is the Kronecker delta, and
\beq
g(r)=\left\{
\begin{array}{lll}
k_0r,~~~R_2<r<R_3,\\
~~\\
\omega\sqrt{\epsilon'\left[f(r)\right]\mu'\left[f(r)\right]}f(r),~~~r<R_2,~r>R_3.
\end{array}
\right.
\eeq
Referring the reader to \cite{Castaldi} for details on the procedure for computing the expansion coefficients $a_n^{(\nu)}$ and $b_n^{(\nu)}$, we limit ourselves to summarize the final results that are instrumental for what follows:
\beq
b_n^{\left(4,5\right)}=i a_n^{\left( 5 \right)},~~~
a_n^{(4)}= i^n+ ia_n^{(5)},~~~
b_n^{(1,2)}= 0,~~~
a_n^{(2)} = a_n^{(1)},
\label{eq:cc1}
\eeq
\beq
a_0^{(1,3,5)}  \sim b_0^{\left( {3} \right)} \sim O\left( {\frac{1}{{\log \Delta _3 }}} \right),
\label{eq:cc2}
\eeq
\beq
a_n^{(1)} \sim O\left( {\frac{{\Delta _3^{\left| n \right|} }}{{\Delta _2^{\left| n \right|} }}} \right),~~~
a_n^{(3)}  \sim b_n^{(3)} \sim O\left( {\Delta _3^{\left| n \right|} } \right),~~~
a_n^{(5)} \sim O\left( {\Delta _3^{2\left| n \right|} } \right), ~~~ n>0,
\label{eq:cc3}
\eeq
where the Landau asymptotic notation $O(\cdot)$ is used, neglecting irrelevant constants and higher-order terms.
From (\ref{eq:cc1})--(\ref{eq:cc3}), it is clear that, letting to zero $\Delta_2$ and $\Delta_3$ while maintaining their ratio {\em finite}, it is possible to render all the expansion coefficients pertaining to the vacuum gap and exterior region {\em vanishingly small}, while keeping the (higher-order) coefficients pertaining to the target region {\em nonzero}. 

\subsection{Slightly mismatched target}
In the new sensor-cloaking scenario of interest, the inner region $r<R_1$ is filled by an absorbing particle (modeling a sensor or a detector), and thus, different from \cite{Castaldi}, is not matched with the surrounding anti-cloak.
We therefore begin by studying {\em perturbatively} the effects of a slightly mismatched electrical permittivity, 
\beq
{\bar \varepsilon}_1  = \left( {1 + \Delta _\varepsilon  } \right)^2 \varepsilon _1, 
\eeq
of the target, with the vanishingly small term $\Delta_{\varepsilon}$ parameterizing this mismatch.  Denoting with an overbar the expansion coefficients in (\ref{eq:BFE}) pertaining to the mismatched configuration, and repeating the procedure in \cite{Castaldi}, with the dependence on $\Delta_{\varepsilon}$ parameterized via a first-order Taylor expansion, we obtain:
\beq
{\bar b}_n^{(1)}=0,~~~{\bar b}_n^{(4,5)}=i {\bar a}_n^{(5)},
~~~{\bar a}_n^{(4)}  = i^n  + {\bar a}_n^{(5)},
\eeq
\beq
{\bar b}_n^{(2)}\sim {\bar a}_n^{(1)} O\left( {\Delta _\varepsilon  } \right),~~~
{\bar a}_n^{(2)}\sim {\bar a}_n^{(1)} \left[ {1 + O\left( {\Delta _\varepsilon  } \right)} \right],
\label{eq:bb}
\eeq
\beq
{\bar a}_n^{(1)} \sim\frac{{a_n^{(1)} }}
{1 + O_n},~~~
\left\{ \begin{array}{l}
 {\bar a}_n^{(3,5)}  \\ 
 {\bar b}_n^{(3)}  \\ 
 \end{array} \right\} \sim \left\{ \begin{array}{l}
 a_n^{(3,5)}  \\ 
 b_n^{(3)}  \\ 
 \end{array} \right\}\left( {\frac{{1 + O_n}}{{1 + O_n}}} \right),
\label{eq:an}
\eeq
where 
\beq
O_n=
\left\{
\begin{array}{lll}
O\left(\Delta_\varepsilon\log\Delta_3\right),~~~n=0,\\
~\\
O\left(\Delta_\varepsilon \Delta_3^{-2|n|}\right),~~~n>0.
\end{array}
\right.
\label{eq:OO}
\eeq
In (\ref{eq:bb}) and (\ref{eq:an}), for a better understanding of the scaling behaviors, the original (i.e., in the absence of mismatch) expansion coefficients have been factored out. It is readily observed that the mismatch affects primarily the field transmitted in the target region, while the field in the vacuum gap and the one scattered in the exterior space maintain the vanishingly small character as in the matched case.  In particular, depending on ratio of the vanishingly small parameters $\Delta_{\varepsilon}$ and $\Delta_3$, the higher-order ${\bar a}_n^{(1)}$ coefficients will exhibit an algebraically decaying behavior.  For instance, assuming $\Delta_{\varepsilon}\sim O(\Delta_3^{2q})$  (with $q>0$) and $\Delta_2\sim O(\Delta_3)$, it results from  (\ref{eq:an}) [with (\ref{eq:cc3})]
\beq
{\bar a}_n^{(1)} \sim O\left( {\Delta _3^{2\left| n \right| - 2q} } \right),\quad n > q.
\label{eq:ho}
\eeq
To sum up, the presence of a slightly mismatched target does not affect the cloaking function, but it may affect the anti-cloaking capability of restoring a field in the target region since, besides the zeroth-order coefficient $a_0^{(1)}$ in (\ref{eq:an}) (which is still logarithmically vanishing as in \cite{Castaldi}), an infinite number of higher-order terms are now generally lost [cf. (\ref{eq:ho})].  For increasing mismatch levels, the anti-cloaking effect is progressively destroyed, and the cloaking effect eventually prevails.

\subsection{Vanishing-gap limit}
We now move on to studying the effects of the vacuum gap $R_2<r<R_3$ which, in the new scenario of interest, is no longer functional to creating a cloaked region, but it constitutes an additional degree of freedom to tailor the cloak/anti-cloak interactions.  While its general effects in the EM response are not easy to parameterize, it is particularly insightful to study the limiting case,
\beq
R_3  = \left( {1 + \Delta _G } \right)R_2,
\label{eq:gap}
\eeq
where the term $\Delta_G$ parameterizes the the vanishingly small gap. Once again, proceeding as in \cite{Castaldi}, and parameterizing as a first-order Taylor expansion the dependence on $\Delta_G$, we obtain for the relevant expansion coefficients in the target and exterior regions:
\beq
a_n^{(1)}
\! \sim\! \left\{\! \begin{array}{l}\!
\displaystyle{\frac{1}{{1 + O\left( {\Delta _G \log \Delta _3 } \right)}}},~~~ n=0, \\
~\\ 
\displaystyle{\frac{{2i^n \gamma ^{\left| n \right|} \varepsilon _1 }}{{\left( {\varepsilon _1  + \varepsilon _0 } \right)}}\!\left({\frac{{\Delta _3 }}{{\Delta _2 }}} \right)^{\!\!\left| n \right|}}\!\!+\! O\!\left( {\Delta _3^{2\left| n \right|} } \right) \!+\!O\!\left({\Delta _G }\right),~~~ n \ne 0, \\ 
 \end{array} \right.
\label{eq:an1G}
\eeq
\beq
a_n^{\left( 5 \right)}  \sim \left\{ \begin{array}{l}
\displaystyle{\frac{{O\left( {\Delta _G \log ^{ - 1} \Delta _3 } \right)}}{{O\left( {\Delta _G } \right) + O\left( {\log ^{ - 1} \Delta _3 } \right)}}},~~~ n = 0, \\ 
~\\ 
 O\left( {\Delta _3^{2|n|} } \right)\left[ {1 + O\left( {\Delta _G } \right)} \right],~~~n \ne 0, \\ 
 \end{array} \right.
\label{eq:an5G}
\eeq
where 
\beq
\gamma=\displaystyle{\frac{\displaystyle{\sqrt {\varepsilon _0 \mu _0} R_4 (R_2  - R_1)}}{\displaystyle{\sqrt {\varepsilon _1 \mu _1 }R_1(R_4  - R_3)}}}.
\eeq

We observe that, in the vanishing-gap limit, the scattered-field coefficients maintain their vanishingly small character [in fact, the zero-th order terms vanish now faster than logarithmically, cf. (\ref{eq:an5G})]. In connection with the field transmitted in the target region [cf. (\ref{eq:an1G})], we point out that, besides the $(n\ne0)$th-order terms [for $\Delta_2\sim O(\Delta_3)$], it is now possible to recover also the zeroth-order term by letting $\Delta_G\rightarrow 0$  faster than $\log^{-1}\Delta_3$.  
In fact, assuming $\Delta_G=0$  (i.e., $R_3=R_2$), and enforcing the continuity of the coordinate transformation in (\ref{eq:CT}), $f(R_2^-=R_2^+)$, we obtain
\beq
\frac{{R_1 \Delta _2 }}{{R_2  + \Delta _2  - R_1 }} = \frac{{R_4 \Delta _3 }}{{R_4  + \Delta _3  - R_3 }},
\eeq
which, for $\Delta_{2,3}\rightarrow 0$, implies 
\beq
\gamma  = \frac{{\sqrt {\varepsilon _0 \mu _0 } }}{{\sqrt {\varepsilon _1 \mu _1 } }}\left( {\frac{{\Delta _2 }}{{\Delta _3 }}} \right).
\eeq
For the (trivial) case of a vacuum target ($\varepsilon_1=\varepsilon_0$, $\mu_1=\mu_0$), this yields $a_n^{(1)}=i^n$  in (\ref{eq:an1G}), i.e., the anti-cloak {\em perfectly compensates} the cloak, restoring the impinging plane-wave, as in the original scenario presented in \cite{Chen}.  For materials other than vacuum, the field transmitted in the target is a distorted version of the incident one.

\subsection{Remarks}
The two cases above (slightly mismatched target and vanishing gap) can readily be combined by substituting (\ref{eq:an1G}) and (\ref{eq:an5G}) in (\ref{eq:an}). From the above results, in the sensor-cloaking scenario of interest, the vanishing-gap configuration appears particularly well suited, since it allows the recovery of the (otherwise logarithmically vanishing) zeroth-order terms of the transmitted field (and thus, in principle, a more effective power absorption), while keeping the scattered field vanishingly small (with a faster decay of the zeroth-order terms). These considerations are expected to hold also in the case of slightly lossy cloak/anti-cloak shells.

\section{Parametric analysis}
\label{Sec:parametric}

\subsection{Generalities and observables}

To complement the analytical results above, we carried out a series of parametric studies, aimed at identifying the ``optimal'' parameter configurations and possible tradeoffs, within and beyond the limits $\Delta_{2,3,G,\varepsilon}\rightarrow 0$, and also in the more realistic case of slightly lossy cloak and anti-cloak. 

In our studies below, we utilize the analytical solution in (\ref{eq:BFE}), and compactly parameterize the EM response of the configurations in terms of the total scattering and absorption cross-sectional widths per unit length \cite{Ruan} (normalized to the vacuum wavelength $\lambda_0$),
\beq
Q_s\!=\!\!\frac{2}{\pi}
\!\sum\limits_{n=-\infty}^\infty
\!\left|{\bar a}_n^{(5)}\!\right|^2\!\!,~~~
Q_a\!\!=\!-\frac{2}{\pi}
\!\sum\limits_{n=-\infty}^\infty
\!\left\{
\!\left|{\bar a}_n^{(5)}\!\right|^2
\!\!+\!
{\mathop{\rm Re}\nolimits}
\!\left[
i^{ -n} {\bar a}_n^{(5)}
\!\right]
\!\right\}\!.
\label{eq:QQ}
\eeq
In the case of slightly lossy cloak and anti-cloak, for which the absorption cross-section $Q_a$ is no longer representative of the power absorption of the target/sensor only, we consider the time-averaged power (per unit length along the $z$-axis) dissipated in the target region,
\beq
P_a  = \frac{{\omega {\mathop{\rm Im}\nolimits} \left[ {{\bar \varepsilon}_1 } \right]}}{2}\int_0^{R_1 } {rdr} \int_0^{2\pi} {d\phi } \left| {{\bf E}\left( {r,\phi } \right)} \right|^2, 
\label{eq:Pa}
\eeq
with ${\bf E}$ denoting the local electric field. 

The configuration under study is characterized by several geometrical and constitutive parameters, which render an {\em exhaustive} optimization computationally prohibitive. 
The designs presented below have been obtained by tweaking a limited number of parameters, until reaching reasonably good performance. Further improvements may be sought, although beyond the interest of the present paper.

\subsection{Representative results}

\begin{figure}
\begin{center}
\includegraphics[width=13.5cm]{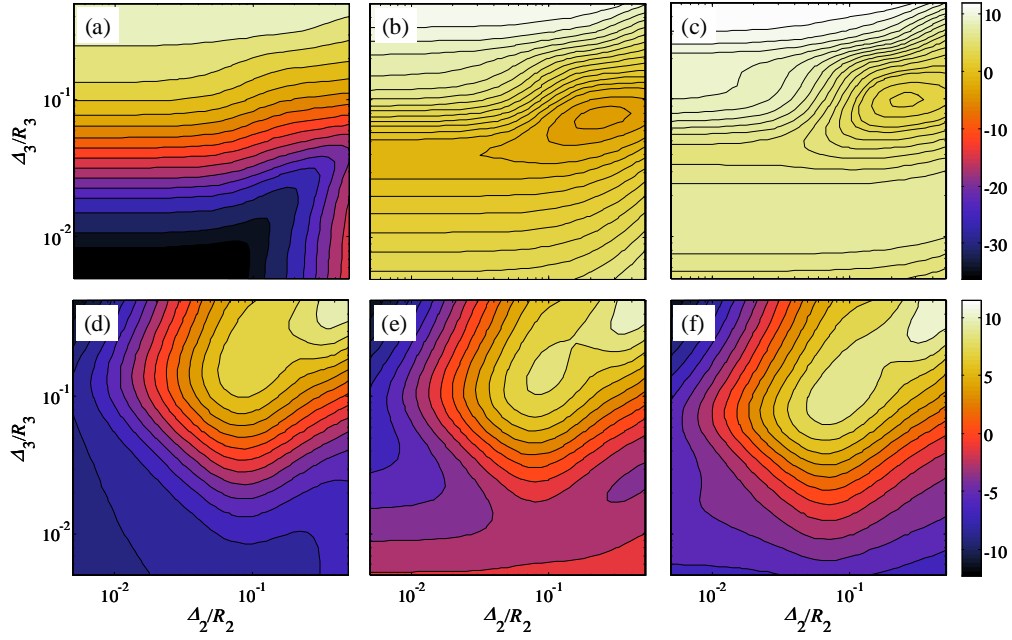}
\end{center}
\caption{(Color online) Geometry as in Fig. \ref{Figure1}, with ${\bar \varepsilon}_1=(4+i0.25)\varepsilon_0$, $\mu_1=\mu_0$, $R_1=\lambda_0/8$, $R_2=\lambda_0/2$, $R_3=(1+\Delta_G)R_2$, $R_4=\lambda_0$, and lossless cloak and anti-cloak. 
(a)--(c) Total scattering cross-sectional
width in (\ref{eq:QQ}) normalized by the reference value in vacuum [cf. (\ref{eq:Qbar})] in dB scale, as a function of
$\Delta_2/R_2$ and $\Delta_3/R_3$, for gap thickness $\Delta_G=0, 1/50, 1/25$, respectively. (d)--(f) Corresponding
total absorption cross-sectional width normalized by the reference value in vacuum.}
\label{Figure2}
\end{figure}

We begin considering the ideal case of lossless cloak and anti-cloak. For a lossy, small dielectric target/sensor with radius $R_1=\lambda_0/8$ and $\varepsilon_1=(4+0.25i)\varepsilon_0$, the plots in Fig. \ref{Figure2}
illustrate the behavior of the observables of interest, as a function of $\Delta_2$ and $\Delta_3$, for $R_2=\lambda_0/2$, $R_3=(1+\Delta_G)R_2$, $R_4=\lambda_0$, and three
representative values of the vacuum gap thickness $\Delta_G$ in (\ref{eq:gap}). More specifically, Figs. \ref{Figure2}(a)--\ref{Figure2}(c)
and \ref{Figure2}(d)--\ref{Figure2}(f) show (in dB scale) the scattering and absorption cross-sectional widths
(\ref{eq:QQ}), respectively, normalized to their values in vacuum (i.e., in the absence of the cloak/anti-cloak
shells), viz.,
\beq
{\bar Q}_s\equiv \frac{Q_s}{Q_s^{(0)}}, ~~~ {\bar Q}_a\equiv \frac{Q_a}{Q_a^{(0)}}.
\label{eq:Qbar}
\eeq
From the scattering responses [Figs. \ref{Figure2}(a)--\ref{Figure2}(c)], we observe the presence of a broad minimum,
whose position depends on the gap thickness $\Delta_G$. The case in the absence of gap [Fig. \ref{Figure2}(a)] turns
out to provide the most substantial scattering reduction (up to over $30$ dB below the reference value in
vacuum); such minimum is obtained for the smallest values of $\Delta_2$ and $\Delta_3$ in the observation range. Much weaker reductions, and different
positions of the minima, are observed in the presence of even a small gap [Figs. \ref{Figure2}(b) and \ref{Figure2}(c)]. In
all three cases, enhancements of the scattering responses (as compared with the reference values in
vacuum) can be observed in certain regions; such {\em superscattering} phenomena \cite{Yang} are typical
of complementary media, but are not of interest in the present investigation.

The absorption responses [Figs. \ref{Figure2}(d)--\ref{Figure2}(f)] exhibit a slightly more complex structure, with the
presence of local minima and maxima, qualitatively similar for the different gap thicknesses. Also
in these cases, some {\em superabsorption} phenomena \cite{Ng} are visible, but they are generally associated with the above
superscattering effects, and therefore not functional in the sensor-cloaking scenario of interest.

\begin{figure}
\begin{center}
\includegraphics[width=10cm]{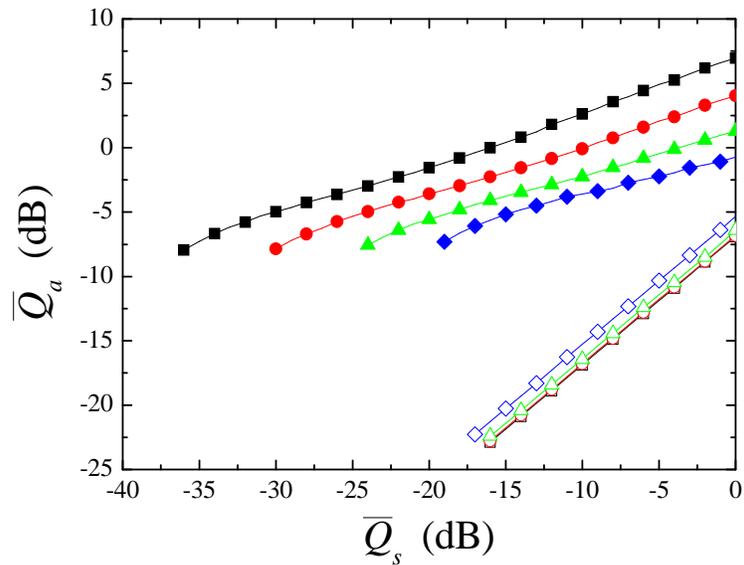}
\end{center}
\caption{(Color online) Tradeoff curves (full markers) pertaining to the lossless cloak/anti-cloak configuration in Fig. \ref{Figure2}, for various values of the target loss levels (squares, circles, triangles, diamonds: $\mbox{Im}[{\bar \varepsilon}_1]=0.25\varepsilon_0$, $0.5\varepsilon_0$, $\varepsilon_0$, $2\varepsilon_0$, respectively). Also shown (empty markers), as references, are the corresponding curves pertaining to an imperfect (lossless) cloak configuration [cf. (\ref{eq:CT1})].}
\label{Figure3}
\end{figure}

\begin{figure}
\begin{center}
\includegraphics[width=9cm]{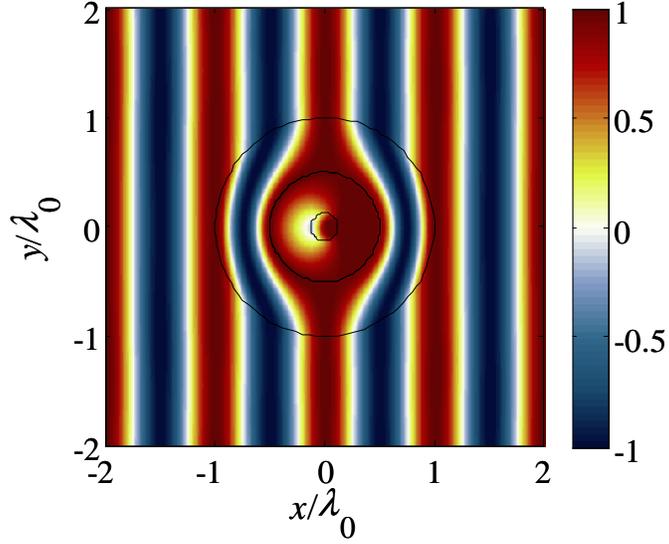}
\end{center}
\caption{(Color online) Geometry and parameters as in Fig. \ref{Figure2}. Field map (real part of magnetic field) pertaining
to a plane-wave-excited lossless cloak/anti-cloak configuration with
$\Delta_G=0$, $\Delta_2/R_2=0.118$, $\Delta_3/R_3=0.03$, featuring ${\bar Q}_s=-20$ dB  and ${\bar Q}_a=-1.56$ dB.}
\label{Figure4}
\end{figure}

\begin{figure}
\begin{center}
\includegraphics[width=9cm]{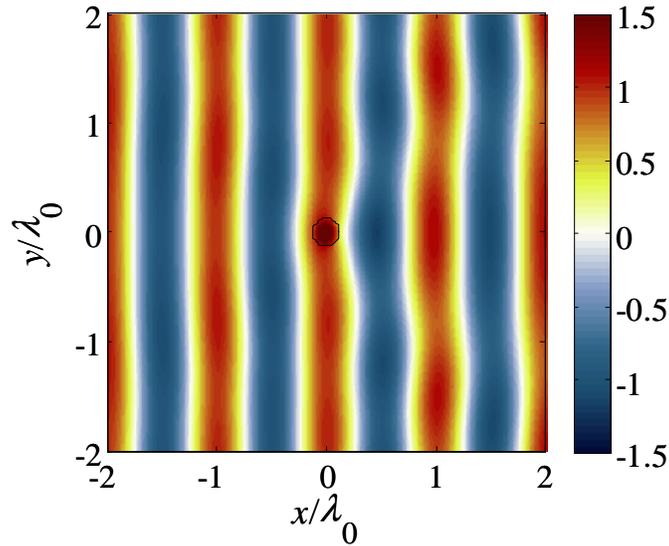}
\end{center}
\caption{(Color online) As in Fig. \ref{Figure4}, but for the target free-standing in vacuum.}
\label{Figure5}
\end{figure}

In agreement with the predictions from the analytical study in Sec. \ref{Sec:analytica}, the potentially more interesting responses
(especially in terms of scattering reductions) are obtained in the absence of the gap.  
However, it can be observed that the maximum scattering reduction generally does not coincide with the maximum absorption, and different parameters choices (e.g., increasing the normalized absorption at the expense of the scattering reduction, and viceversa) are possible, as summarized in the tradeoff curve (full square markers) in Fig. \ref{Figure3}. Such curve, extracted from Figs. \ref{Figure2}(a) and \ref{Figure2}(d), for a given value of the
(normalized) scattering response yields the largest (normalized) absorption response attainable. In
particular, it illustrates how, in the absence of the gap and varying $\Delta_2$ and $\Delta_3$, it is possible to span
the entire range of cloak/anti-cloak interactions, going from a regime featuring weak scattering and
weak absorption (i.e., cloak-prevailing) to one featuring scattering and absorption levels comparable
with those in vacuum (i.e., anti-cloak-prevailing). However, the slope of the curve is such that, in an intermediate regime, one can attain significant scattering reductions (e.g., -20 dB) while maintaining sensible absorption levels (e.g., -1.56dB). 
Figure \ref{Figure4} shows a representative field map, from which the very weak overall
scattering, and yet the power-coupling capabilities, are fairly visible; as a reference, the case of the
target/sensor free-standing in vacuum is shown in Fig. \ref{Figure5}.

For comparison purposes, Fig. \ref{Figure3} also shows (full markers) the tradeoff curves pertaining to different levels of 
target losses ($\mbox{Im}[{\bar \varepsilon}_1]$), from which one notes that loss (and hence mismatch) increases result in a performance degradation (i.e., lower normalized absorption for a given scattering reduction), in accord with our theoretical predictions in Sec. \ref{Sec:analytica}. 

One might argue that a similar trend (i.e., scattering reduction at the expenses of absorption) is also qualitatively exhibited by an imperfect cloak {\em alone}, and might therefore wonder to what extent the anti-cloak is strictly necessary and functional. In order to address this issue, Fig. \ref{Figure3} also shows (empty markers) the reference curves pertaining to a TO-based imperfect cloak characterized by the standard (linear) coordinate transformation 
\beq
r' =f(r)= \left\{ \begin{array}{l}
r,~~ r < R_1 ,~~r > R_4 , \\ 
~\\
 R_4 \left( \displaystyle{{\frac{{r - R_1  + \Delta_1 }}{{R_4  - R_1  + \Delta_1}}}} \right),~~ R_1  < r < R_4, \\ 
\end{array} \right.
\label{eq:CT1}
\eeq
and the same total thickness ($R_4-R_1$) as the above cloak/anti-cloak configuration. Note that, in this case, the resulting  tradeoff curves are obtained by simply scanning the only parameter ($\Delta_1$) available. One can observe that, in spite of the anticipated qualitatively similar trends, the tradeoff curves pertaining to the imperfect cloak alone are {\em considerably steeper}, and therefore yield acceptable scattering reductions only at the expenses of comparable reductions in the absorption.

\begin{figure}
\begin{center}
\includegraphics[width=13.5cm]{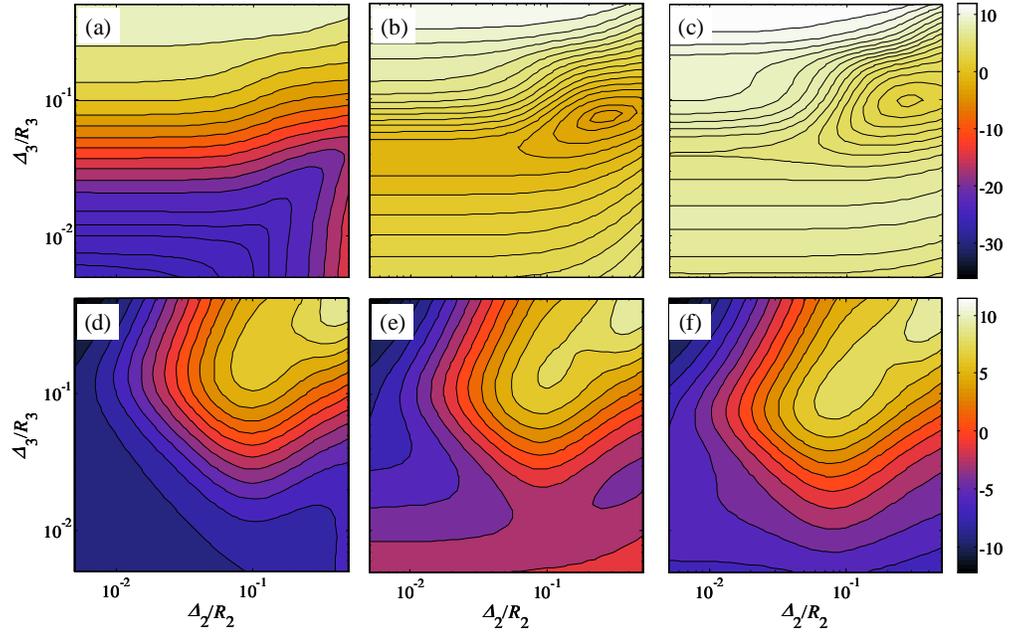}
\end{center}
\caption{(Color online) As in Fig. \ref{Figure2}, but for a lossy cloak (loss-tangent=0.001) and anti-cloak (loss-tangent=0.01) configuration. In (d)--(f),
the time-averaged dissipative power $P_a$ in (\ref{eq:Pa}) (normalized to its free space value $P_a^{(0)}$) is
considered.}
\label{Figure6}
\end{figure}

\begin{figure}
\begin{center}
\includegraphics[width=10cm]{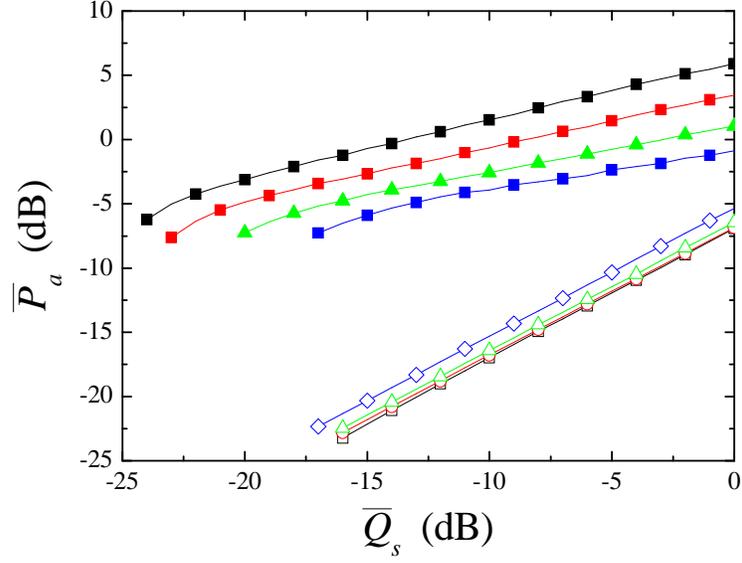}
\end{center}
\caption{(Color online) As in Fig. \ref{Figure3}, but for a lossy cloak (loss-tangent=0.001) and anti-cloak (loss-tangent=0.01) configuration.}
\label{Figure7}
\end{figure}

\begin{figure}
\begin{center}
\includegraphics[width=9cm]{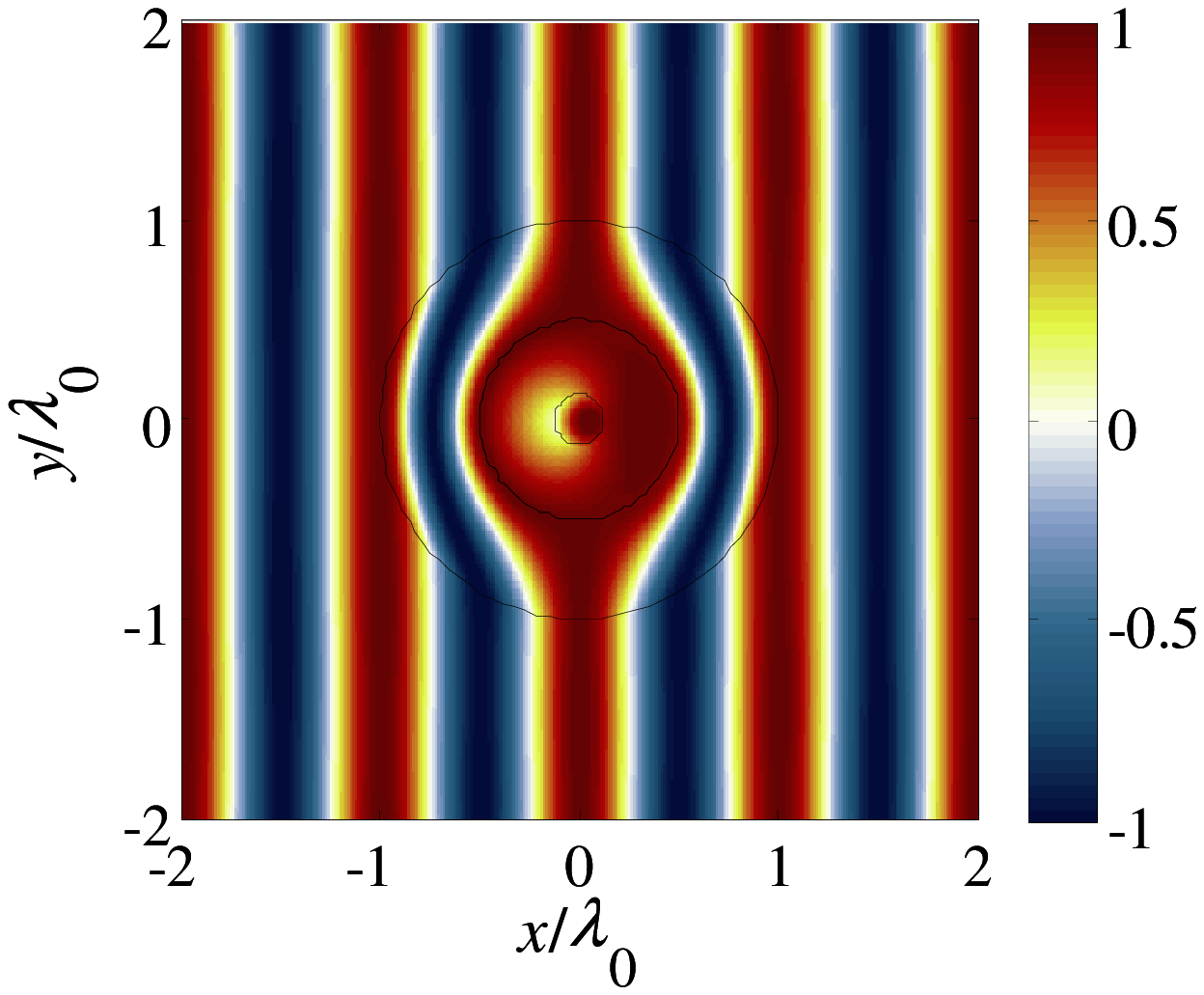}
\end{center}
\caption{(Color online) As in Fig. \ref{Figure4}, but for a slightly lossy cloak (loss-tangent=0.001) and anti-cloak (loss-tangent=0.01) configuration with $\Delta_G = 0$, $\Delta_2/R_2= 0.158$, $\Delta_3/R_3=0.028$, featuring ${\bar Q}_s=-20$ dB and ${\bar P}_a\equiv P_a/P_a^{(0)}=-3.13$ dB.}
\label{Figure8}
\end{figure}

We also considered more realistic cases involving {\em lossy} cloak and anti-cloak shells. More specifically, for the (DPS) cloak shell, we assumed a loss-tangent of 0.001, while for the (DNG) anti-cloak we considered a higher value (0.01).
The corresponding scattering and absorption normalized parameters are shown in Fig. \ref{Figure6}, with the time-averaged dissipative power in (\ref{eq:Pa}) replacing the absorption cross-sectional width in
(\ref{eq:QQ}). As compared with the lossless case above, qualitatively similar results are observed, with an
expectable loss-induced moderate deterioration in the performance. Again, the case in the absence
of the gap [Figs. \ref{Figure6}(a) and \ref{Figure6}(d)] turns out to provide the more interesting responses. 
Figure \ref{Figure7} shows the corresponding tradeoff curves, from which, by comparison with the lossless case in Fig. \ref{Figure3}, one observes a reduced dynamics in the scattering reduction and a slight increase in the slope.
Figure \ref{Figure8} shows a reference field map, for a configuration featuring a scattering reduction of 20 dB at the expense of an absorption reduction of $\sim 3$dB, which qualitatively resembles the lossless case above. 

\begin{figure}
\begin{center}
\includegraphics[width=10cm]{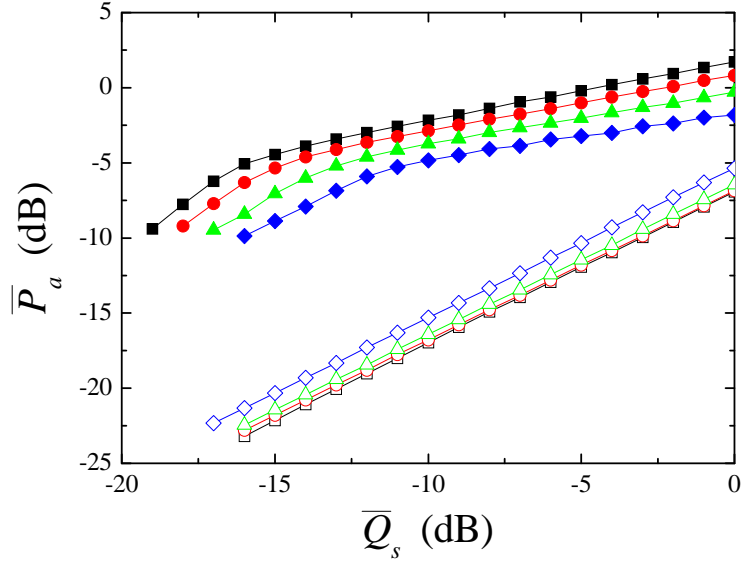}
\end{center}
\caption{(Color online) As in Fig. \ref{Figure3}, but for increased loss level (loss-tangent=0.1) in the anti-cloak.}
\label{Figure9}
\end{figure}

\begin{figure}
\begin{center}
\includegraphics[width=9cm]{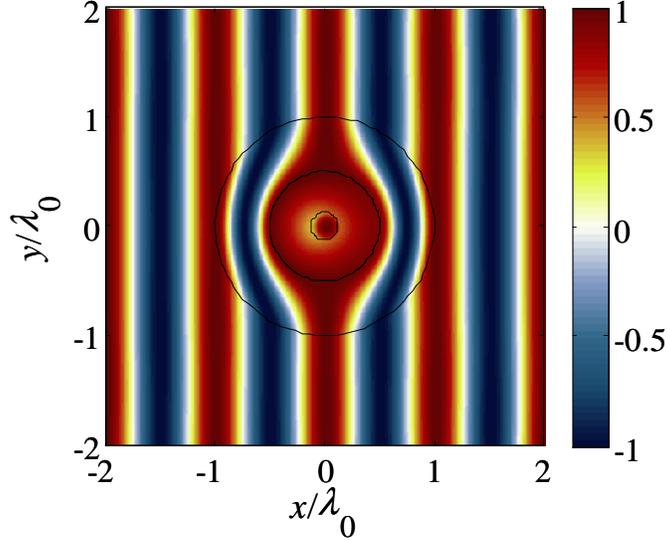}
\end{center}
\caption{(Color online) As in Fig. \ref{Figure4}, but for a lossy cloak (loss-tangent=0.001) and anti-cloak (loss-tangent=0.1) configuration with $\Delta_G = 0$, $\Delta_2/R_2= 0.315$, $\Delta_3/R_3=0.037$, featuring ${\bar Q}_s=-15$ dB and ${\bar P}_a=-4.44$ dB.}
\label{Figure10}
\end{figure}

Finally, Fig. \ref{Figure9} shows the effects in the tradeoff curves of a further increase of the loss-tangent (0.1) in the (DNG) anti-cloak. Again, further compression in the ${\bar Q_s}$ dynamics and increase in the slope are observed. Nevertheless, in spite of the relatively high loss level, in the best-performing case ($\mbox{Im}[{\bar \varepsilon}_1]=0.25\varepsilon_0$), one can still achieve a scattering reduction of nearly 15dB at the expense of a reduction of $\sim 4$ dB in the absorption. The corresponding field map is shown in Fig. \ref{Figure10}.

\section{Implementation issues}
\label{Sec:implementation}
Although the present prototype study is mainly aimed at providing a proof of principle of the cloak/anti-cloak-based sensor-invisibility effect, it is still important to address the related implementation issues. In particular, for the proposed configuration, the major technological challenges associated with the practical realization of the involved transformation media are related to the spatial inhomogeneity (with extreme values) of all constitutive-parameter components, their general magnetic character, and, most of all, the DNG character of the anti-cloak (in view of the assumed DPS target). 

The first two issues are inherent of TO-based cloaks, and can be addressed via rigorous or approximate approaches. In particular, higher-order coordinate transformations have been proposed in order to obtain a perfect cloak with {\em spatially invariant} axial material parameters \cite{Luo}. {\em Non-magnetic} implementations have also been proposed, based on approximate reductions of the constitutive parameters that preserve the ray trajectories inside the cloak shell (see, e.g., \cite{Cummer,Cai2,Gallina1,Zhang}) as well as on mapping from a {\em nearly transparent}, anisotropic and spatially inhomogeneous fictitious domain \cite{Castaldi2}.

On the other hand, the last issue (DNG character) is specifically tied with the anti-cloak concept \cite{Chen}, although it may be transferred, in principle, to the target \cite{Castaldi}. Nevertheless, we have shown \cite{Gallina} in some preliminary studies on square-type configurations (similar to that originally proposed in \cite{Rahm}) the intriguing possibility of achieving approximate anti-cloaking effects using DPS media only. Extension of these preliminary results to the sensor-cloaking scenario is therefore a key issue, and is currently being pursued.

To sum up, a candidate practical implementation may be envisioned as based on the above approximate DPS anti-cloak in conjunction with a suitable parameter reduction approach \cite{Luo,Cummer,Cai2,Gallina1,Zhang,Castaldi2}. The study of such a configuration, which would generally require non-trivial modifications in our analytical framework or a fully-numerical approach, is beyond the scope of the present prototype study, and will be the subject of forthcoming publications.

\section{Conclusions and perspectives}
\label{Sec:conclusions}
In this paper, we have reported our study of cloak/anti-cloak interactions in 2-D cylindrical
geometries, aimed at exploring the potential of such mechanisms in a sensor-cloaking scenario. Via
analytical and parametric studies, we have shown that it is possible to tailor the cloak/anti-cloak
interactions so as to strongly reduce the scattering response of a slightly lossy cylindrical target
(mimicking a ``sensor'') while still allowing effective dissipative power coupling, with a controllable
tradeoff. Such effects are not destroyed by the presence of slight/moderate losses in the cloak/anti-cloak.

These results lay the foundations for a possible TO-based route to sensor-cloaking, alternative to the
insofar proposed approaches \cite{Alu2,Ruan,Greenleaf}. Within this framework, a comparative study of the various approaches is in order. For instance, one difference that may be expected between our approach (which generally requires DNG resonant metamaterials) and the scattering-cancellation approach in \cite{Alu2} (which is based on inherently penetrable and non-resonant materials) is the presence in our case of a ``delay'' effect (typical of DNG-based devices) due to the buildup-time of the resonant field in the anti-cloak. 

Current and future studies are also aimed at removing or mitigating the major technological challenges associated with the practical realization of the involved transformation media (see Sec. \ref{Sec:implementation}), as well as at the extension to the full 3-D (vector) scenario.

\section*{Acknowledgments} 
A. Al\`u was partially supported by the National Science Foundation (NSF) CAREER Grant No. ECCS-0953311.


\newpage

%

%

%

%

%

%

%

%

%

%

\end{document}